\documentstyle[aps,pra,prabib,amsmath,amssymb,amsfonts,epsfig,preprint]{revtex}

\newcommand{\cen}[1]{\begin{center}#1\end{center}}
 
\newcommand{\eqname}[1]{\label{eq:#1}}
\newcommand{\bgar}{\begin{eqnarray}}
\newcommand{\enar}[1]{\label{eq:#1}\end{eqnarray}}
\newcommand{\valass}[1]{\left|#1\right|}
\newcommand{\norme}[1]{\left\|#1\right\|}
\newcommand{\trace}[1]{\textrm{Tr}\left[#1\right]}

\newcommand{\braket}[2]{ \langle #1 | #2 \rangle }
\newcommand{\braopket}[3]{ \left\langle #1 \right| #2 \left| #3 \right\rangle }
\newcommand{\ketbra}[2]{ | #1 \left\rangle \right\langle #2 |}

\newcommand{\derivexp}[2]{\frac{\partial #1}{\partial #2}}

\newcommand{\mean}[1]{\overline{#1}}
\newcommand{\expect}[1]{\left\langle #1 \right\rangle}

\newcommand{\x}{ {\bf x}}

\newcommand{\eq}[1]{(\ref{eq:#1})}
\newcommand{\al}[1]{^{(#1)}}
\newcommand{\Psihd}{\hat\Psi^\dagger}

\newcommand{\Psih}{\hat\Psi}

\newcommand{\Hamilt}{{\mathcal H}}
\newcommand{\ahd}{\hat a^\dagger}
\newcommand{\ah}{\hat a}

\begin{document}
\tightenlines
\setlength{\parindent}{0cm}
\setlength{\parskip}{0.1cm}

\title{An exact stochastic field method for the interacting Bose gas
at thermal equilibrium}
\author{I. Carusotto\footnote{E-mail: Iacopo.Carusotto@lkb.ens.fr} and
Y. Castin}
\address{Laboratoire Kastler Brossel de l'\'Ecole Normale Sup\'erieure \\ 24 rue
Lhomond, 75231 Paris Cedex 05, France}
\date{\today}

\maketitle

\begin{center}{\sc \bf ABSTRACT}
\end{center}

We present a new exact method to numerically compute the thermodynamical properties of an
interacting Bose gas in the canonical ensemble. As in our previous paper (Phys. Rev. A
{\bf 63} 023606 (2001) ), we write the density operator $\rho$ as an average of
Hartree dyadics $\ketbra{N:\phi_1}{N:\phi_2}$ and we find stochastic evolution
equations for the wave functions $\phi_{1,2}$ such that the exact imaginary-time
evolution of $\rho$ is recovered after average over noise. In this
way, the thermal equilibrium density operator can be obtained for any
temperature $T$.
The method is then applied to study the thermodynamical
properties of a homogeneous one-dimensional $N$-boson system: 
although Bose-Einstein
condensation can not occur in the thermodynamical limit, 
a macroscopic
occupation of the lowest mode of a finite system is observed at sufficiently low temperatures.
If $k_B T \gg \mu$, the main effect of interactions is to
suppress density fluctuations and to reduce their correlation length.
Different effects such as a spatial antibunching of the atoms are
predicted for the opposite $k_B T\leq \mu$ regime.
Our exact stochastic calculations have been compared to existing
approximate theories.


\section{Introduction}

The recent advances in the production and manipulation of ultra-cold
atomic samples have opened the way to
the investigation of the thermodynamical and dynamical properties of
interacting Bose gases in different geometries and dimensionalities~\cite{1DBEC,1DFluctPhase,2Dexp};
in three dimensional traps, Bose-Einstein condensation was realized a 
few years ago~\cite{BEC3D} and now there is a great deal of
experimental interest in lower-dimensional systems, for which a
remarkably different physics is predicted~\cite{Lieb,2DColl,1DOlsh,Shlyap1D,2D,Ketter2D}.

In presence of a condensate, i.e. when the coherence of the Bose field
extends throughout the whole sample, a mean-field theory can be
applied, which completely neglects the non-condensed atoms and
describes the system in terms of the classical field corresponding to
the condensate wave function~\cite{StringReview}.
In three dimensions, such an approach is clearly valid only when a
small non-condensed fraction is present, which implies a
low temperature and weak interaction regime; for
low-dimensional systems a further constraint has to be imposed on the
spatial size of the system, which has to be smaller than the finite coherence
length of the field.
Within mean-field theory, weak thermal excitations on top of a well
populated condensate can be taken into account by means of the
Bogoliubov approach which describes the low-lying part of
the many-body spectrum as a collection of uncoupled harmonic
oscillators~\cite{Bogo,Huang,CCTCollFrance,YvanBogo}.
Clearly, this approximated theory is valid only at relatively low
temperatures, when the density of thermal excitations is small enough
for their mutual interaction to be neglected.

Exact results for the thermodynamical properties of interacting Bose
gases at any temperature can in principle be obtained using Quantum Monte Carlo
techniques~\cite{QMC}, but such an approach is subjected to strict
limitations in its applicability domain: first of all, it requires
that the matrix elements of the Hamiltonian are real in the position
representation; this condition excludes rotating systems such as the
ones used to study vortices.
The privileged role attributed to the position
representation makes some observables hard to be sampled, e.g. the
momentum distribution function or the occupation number statistics of
a given mode.
Finally, the thermodynamical calculations performed with the
quantum Monte Carlo method can not be prolongated to real time so 
to determine the dynamical properties of the system.

In the present paper we present a new exact and numerically tractable
approach to the calculation of thermodynamical properties of an 
interacting Bose gas
in the canonical ensemble. This computational scheme is not restricted
to Hamiltonians with real matrix elements in real space and gives
access to all observables of the system.
Also, this computational scheme can be
easily integrated with our reformulation of the dynamical
evolution~\cite{GPstoch} and therefore used to describe the response
of finite-temperature samples to perturbations away from equilibrium.
The method is based on the stochastic evolution of Hartree dyadics
$\sigma=\ketbra{N:\phi_1}{N:\phi_2}$ where $\phi_{1,2}$ are
non-normalized wave functions. 
In our previous paper~\cite{GPstoch} we showed that it
is possible to choose the
stochastic evolution of $\phi_{1,2}$ in such a way that
 the average evolution of $\sigma$ over the stochastic noise exactly
reproduces the full many-body time-evolution. Here, we
show that similar arguments can be used to find a stochastic evolution
which exactly reproduces the imaginary-time evolution of the Bose gas 
and therefore
can be used to determine the thermodynamical properties of the
interacting $N$-boson system in the canonical ensemble.
Differently from previous theories such as Positive-$P$
representation, our scheme is not subjected to instability problems~\cite{Divergences}.

After the general discussion on the method, we shall report its first
application to a homogeneous one-dimensional interacting Bose gas:
such a model is expected to accurately describe
the physics of Bose condensates tightly confined in the transverse
dimensions so to create an effective one-dimensional
geometry~\cite{1DBEC}.

The outline of the paper is the following: in Sec.\ref{sec:Stoch} we
present the general theory underlying the method and we discuss the
stochastic equations for the Hartree wave functions $\phi_{1,2}$
to be used for the study of dynamics and thermodynamics of the interacting
Bose gas.

In Sec.\ref{sec:Simul}, we give details on the algorithm used in
numerical simulations in order to enhance the efficiency of Monte
Carlo calculations; in particular, a large reduction in the
statistical noise can be obtained by means of an importance sampling
with an appropriate {\em a priori} distribution function.

In Sec.\ref{sec:1D}, we shall apply the method to the
thermodynamical properties of a homogeneous one-dimensional
interacting Bose gas in the canonical ensemble and we shall compare
the results of our stochastic calculations with existing approximate 
theories.

Conclusions are finally given in Sec.\ref{sec:Conclu}

\section{The stochastic wave function approach}
\label{sec:Stoch}
The Hamiltonian of a trapped interacting Bose gas in $D$ dimensions 
can be written in a second-quantized form as
\begin{equation}\label{eq:Hamilt}
{\mathcal H}=\int\!d^D\x\,\hat\Psi^\dagger(\x) h_0 \hat\Psi(\x)+\frac{1}{2}
\int\!\int\!d^D\x\,d^D\x'\,\hat\Psi^\dagger(\x)\hat\Psi^\dagger(\x')
V_{\rm int}(\x-\x')\hat\Psi(\x')\hat\Psi(\x)
\end{equation}
where $h_0=-\frac{\hbar^2}{2m}\nabla^2+V_{\rm ext}(\x)$ is the 
single particle Hamiltonian in the external confining potential 
$V_{\rm ext}(\x)$. Interactions are modeled by the two-body potential
$V_{\rm int}(\x-\x')$ which is assumed to be repulsive.
The field operators $\Psi(\x)$ satisfy Bose commutation relations 
$[\Psih(\x),\Psihd(\x')]=\delta(\x-\x')$.

Although the exact dynamics of the
many-body system could in principle be obtained from the quantum
mechanical equation of motion for the $N$-body density operator $\rho$
\begin{equation}
\derivexp{\rho(t)}{t}=\frac{1}{i\hbar}\left[\Hamilt,\rho(t)\right], 
\eqname{MotEqRho}
\end{equation}
any explicit calculation results impossible for the experimentally
relevant multi-mode systems because of the huge
dimensionality of the corresponding $N$-body Hilbert space.

In this section we shall discuss a reformulation of the bosonic
many-body problem in terms of stochastic wave functions.
As we have shown in~\cite{GPstoch}, the $N$-body density operator of the
system can always be written as a statistical average of Hartree operators
$\ketbra{N:\phi_1}{N:\phi_2}$ in which all $N$ atoms share the same,
not necessarily normalized
wave function (Sec.\ref{sec:Ansatz}).
In the same paper, we have shown that the exact time-evolution of the
full many-body system can be reproduced by an appropriately chosen
stochastic evolution of the pair of single-particle wave functions
 $\phi_{1,2}$ (Sec.\ref{sec:Dynam}).
The central theoretical result of the present paper is presented in
Sec.\ref{sec:Thermo}: a similar stochastic evolution of the single
particle wave functions $\phi_{1,2}$ is able to reproduce the
imaginary-time evolution:
\begin{equation}
\derivexp{\rho(\tau)}{\tau}=-\frac{1}{2}\left\{\Hamilt,\rho(\tau)\right\}=-\frac{1}{2}\left(\Hamilt\,\rho(\tau)+\rho(\tau)\,\Hamilt\right)
\eqname{MotEqRhoImag}
\end{equation}
of the many body system with initial condition $\rho(\tau=0)=\textbf{Id}$; this gives access to the thermodynamical
properties of the interacting Bose gas in the canonical ensemble at
any desired temperature $T$ as the solution of \eq{MotEqRhoImag} after
an evolution ``time'' $\tau=1/k_B T$ is $\rho(\tau=1/k_B T)=e^{-\Hamilt/k_B T}$. 

\subsection{The Fock state Hartree dyadic ansatz}
\label{sec:Ansatz}

As in our precedent paper~\cite{GPstoch}, we consider the $N$-body Fock state
Hartree ansatz
\begin{equation}\label{eq:Ansatz}
\sigma=|N:\phi_1\left\rangle \right\langle N:\phi_2|\quad
\end{equation}
in which the two Hartree wave functions $\phi_{1,2}$ in the bra and
the ket are in general different and not normalized.

Given the remarkable identity~\cite{GPstoch} ($\int_1\!{\mathcal D}\phi$
denotes the functional integration over the set of wave functions of unit norm $\|\phi\|=1$)
\begin{equation}
{\bf Id}=\int_1\!{\mathcal D}\phi\,\ketbra{N:\phi}{N:\phi},
\eqname{YvanTh}
\end{equation}
any $N$-body density operator $\rho$ can be expanded over the family of dyadics \eq{Ansatz}
\begin{equation}
\rho=\int\!\int\!{\mathcal D}\phi_1\,{\mathcal D}\phi_2\,{\mathcal P}(\phi_1,\phi_2)\,\ketbra{N:\phi_1}{N:\phi_2}
\eqname{Expansion}
\end{equation}
with a positive probability distribution ${\mathcal P}(\phi_1,\phi_2)$. Note that this expansion is by no means unique and several
distinct  ${\mathcal P}(\phi_1,\phi_2)$ can be found which correspond
to the same physical density operator $\rho$.

From the expansion \eq{Expansion}, the mean values of the observables
can be written as mean values of combinations of $\phi_1$ and $\phi_2$ over
the probability distribution ${\mathcal P}(\phi_1,\phi_2)$.
As simple examples, the one-body density matrix $\rho\al{1}(\x,\x')$ can be written as
\begin{equation}
\rho\al{1}(\x,\x')=\expect{\Psihd(\x')\Psih(\x)}=N\mean{\phi_1(\x)\phi_2^*(\x')\braket{\phi_2}{\phi_1}^{N-1}}
\eqname{r1}
\end{equation}
and the second-order correlation function $C\al{2}(\x,\x')$ as
\begin{multline}
C\al{2}(\x,\x')=\expect{\Psihd(\x)\Psihd(\x')\Psih(\x')\Psih(\x)}=\\ =N(N-1)\;\mean{\phi_1(\x)\phi_1(\x')\phi_2^*(\x')\phi_2^*(\x)\braket{\phi_2}{\phi_1}^{N-2}}.
\eqname{C2}
\end{multline}

\subsection{Real-time evolution: dynamics}
\label{sec:Dynam}

The central result of the paper~\cite{GPstoch} was the proof of the
possibility of reproducing the exact time evolution \eq{MotEqRho} of
$\rho$ by means of a Ito stochastic evolution for $\phi_{1,2}$
\begin{equation}\label{eq:HartreeEvStoch2}
\phi_\alpha(t+d t)=\phi_\alpha(t)+F_\alpha\; d t + d B_\alpha 
\qquad (\alpha=1,2);
\end{equation}
in all the paper, the noise term $dB_\alpha$ is treated in the
standard Ito formalism~\cite{StochMeth}: it has a zero mean $\mean{dB_\alpha}$
and $dB_\alpha^2\propto dt$; the deterministic contribution is given by
the force term $F_\alpha\,dt$.

In the infinitesimal time interval $[t,t+dt]$, the two wave functions
$\phi_{1}(\x)$ and $\phi_2(\x)$ are assumed to evolve according to
Ito stochastic differential equations \eq{HartreeEvStoch2}.
After $dt$, the dyadic $\sigma(t)=\ketbra{N:\phi_1}{N:\phi_2}$ has therefore evolved into
\begin{equation}
\eqname{sigmadt}
\sigma(t+dt)=\ketbra{N:\phi_1+d\phi_1}{N:\phi_2+d\phi_2}
\end{equation}
where $d\phi_{1,2}$ are defined according to \eq{HartreeEvStoch2}.
Imposing that this evolution reproduces in the average the exact many-body evolution
\begin{equation}
\mean{\sigma(t+dt)}=\sigma(t)+\frac{dt}{i\hbar}(\Hamilt\sigma(t)-\sigma(t)\Hamilt)
\eqname{sigmaEx}
\end{equation}
gives a set of constraints on the explicit form of the stochastic
differential equations \eq{HartreeEvStoch2}.

As we have shown in~\cite{GPstoch}, many stochastic evolutions can be
found which all satisfy the exactness constraints: their stability and 
statistical properties are however significantly different.
In the following we shall limit our attention to the so-called {\em
simple scheme} minimizing the growth of the statistical variance of $\sigma$: thanks to its stability, this scheme has in
fact been found to be the most efficient for practical simulations.

Explicitly, the stochastic evolution of the wave functions $\phi_{1,2}$ reads
\begin{multline}\eqname{stoch1}
d\phi_\alpha=\frac{1}{i\hbar}\left[-\frac{\hbar^2\nabla^2}{2m}+V_{ext}(\x)+\frac{(N-1)}{\norme{\phi_\alpha}^2}\int\!d\x'\,V_{\rm
int}(\x-\x')\valass{\phi_\alpha(\x')}^2+\right.\\
\left.-\frac{(N-1)}{2}\frac{\braopket{\phi_\alpha\phi_\alpha}{V_{\rm int}}{\phi_\alpha\phi_\alpha}}{\norme{\phi_\alpha}^4}\right]\phi_\alpha\,dt+dB_\alpha;
\end{multline}
as expected, the deterministic term closely resembles the
Gross-Pitaevskii evolution of mean-field theory; the noises $dB_1$
and $dB_2$ are statistically independent and are determined
by the correlation properties
\begin{equation}
\eqname{stoch2}
dB_\alpha(\x)\,dB_\alpha(\x')=\frac{dt}{i\hbar}{\mathcal
Q}_\alpha^{\x} {\mathcal Q}^{\x'}_\alpha\left[V_{\rm int}(\x-\x')\phi_\alpha(\x)\phi_\alpha(\x')\right], \quad\quad (\alpha=1,2)
\end{equation}
where ${\mathcal Q}^{\x}_\alpha$ is the projector onto the subspace
orthogonal to $\phi_\alpha(\x)$. 
Thanks to the fact that the deterministic term is norm-preserving, a
mathematically well-defined solution exist for all
times~\cite{GPstoch}
 so that the simulation does not suffer from the divergence problems
encountered by previous stochastic approaches such as Gardiner's
Positive-$P$ distribution~\cite{Divergences,QuantumNoise,WallsMilb,PositiveP}.
The statistical error, although
exponentially growing with time proportionally to $\exp[2N
V_{\rm int}(0)t/\hbar]$, is always finite
\footnote{
This result holds in the case of repulsive interaction
potentials with a positive Fourier transform~\cite{GPstoch}, for which
$V_{\rm int}({\bf 0})\geq |V_{\rm int}(\x)|$. This means, in particular, that
the statistical error only vanishes in the absence of interactions.
}
: this guarantees that reliable physical
results can be obtained by means of a Monte Carlo simulation of the
stochastic differential equations; the norm-preserving character of
the deterministic force ensures that no intrinsically spiking trajectories~\cite{Divergences}
can occur in the numerical calculations~. 
This fundamental fact has been verified in Monte Carlo simulations of
simple systems.

Since any $N$-body density operator $\rho(t)$ can be
expanded in Hartree dyadics \eq{Ansatz}, the time evolution of the
many body system can be reproduced by the stochastic approach starting
from any initial condition.
In~\cite{GPstoch} we used as initial state a fully coherent mean-field state 
$\rho=\ketbra{N:\phi}{N:\phi}$; in order to be able to get information
on the full dynamics keeping track of all quantum effects, a more
physical initial state has to be chosen.
In the following part of the paper we shall show how a related stochastic wave
function approach can be found which is able to sample the thermal
equilibrium density operator $\rho_{eq}(\beta)$ for a many-body system
of $N$ interacting bosons at a given $\beta=1/k_B
T$ in the canonical ensemble.
A simulation of the dynamics of the many body system at finite
temperature $\beta$ can be then performed on the base of
\eq{HartreeEvStoch2} using $\rho_{eq}(\beta)$ as the initial state.

\subsection{Imaginary-time evolution: thermodynamics}
\label{sec:Thermo}
It is a well-known fact of quantum statistical mechanics~\cite{Huang} that the
unnormalized thermal equilibrium density operator $\rho_{eq}$ at a given
temperature $T$ in the canonical ensemble can be written as
\begin{equation}
\rho_{eq}(\beta)=e^{-\beta\Hamilt}
\eqname{EquilRho}
\end{equation}
with $\beta=1/k_B T$.
Starting from the infinite temperature equilibrium state
\begin{equation}\eqname{IdentityMatrix}
\rho_{eq}(\beta=0)={\bf Id},
\end{equation}
the density operator at any given $\beta$ can be obtained by means of
the imaginary-time evolution
\begin{equation}\label{eq:ExactEvImT}
\frac{d\rho_{eq}(\tau)}{d\tau}=-\frac{1}{2}\left\{{\mathcal
H},\rho_{eq}(\tau)\right\}=-\frac{1}{2}\left({\mathcal
H}\,\rho_{eq}(\tau)+\rho_{eq}(\tau)\,{\mathcal H}\right)
\end{equation}
during a ``time'' $\beta$.
The mean values of the observables are then easily obtained from the
equilibrium density operator
\begin{equation}\label{eq:ObservMean}
\langle {\hat O}\rangle=\frac{1}{\trace{\rho_{eq}}}\textrm{Tr} [\rho_{eq}{\hat O}].
\end{equation}

Given the similarity of the real- and imaginary-time evolution equations \eq{MotEqRho} and
\eq{ExactEvImT}, the same arguments which led to 
(\ref{eq:stoch1},\ref{eq:stoch2})
can be generalized to the imaginary-time evolution: the stochastic equations for the wave functions $\phi_{1,2}$ within the
simple scheme now read
\begin{multline}
\eqname{EvolImagT}
d\phi_\alpha=-\frac{d\tau}{2}\left[-\frac{\hbar^2\nabla^2}{2m}+V_{ext}(\x)+\frac{(N-1)}{\|\phi_\alpha\|^2}\int\!d\x'\,V_{\rm
int}(\x-\x')\valass{\phi_\alpha(\x')}^2+\right.\\
\left.-\frac{(N-1)}{2}\frac{\braopket{\phi_\alpha\phi_\alpha}{V_{\rm int}}{\phi_\alpha\phi_\alpha}}{\norme{\phi_\alpha}^4}\right]\phi_\alpha+dB_\alpha;
\end{multline}
and the correlation functions of the statistically independent noise terms $dB_{1,2}$ are fixed by
\begin{equation}
\eqname{NoiseImagT}
dB_\alpha(\x)\,dB_\alpha(\x')=-\frac{d\tau}{2}{\mathcal Q}_\alpha^{\x} {\mathcal
Q}^{\x'}_\alpha\left[V_{\rm int}(\x-\x')\phi_\alpha(\x)\phi_\alpha(\x')\right];
\end{equation}
it is apparent how these equations are obtained from the real-time
ones by means of the simple substitution 
$dt\rightarrow -i\hbar\frac{d\tau}{2}$.

As the starting point of the imaginary-time evolution, one has to
choose a representation of the identity operator describing the
$T=\infty$ equilibrium state; even if not the most efficient one, the 
simplest choice is clearly the one dictated
by \eq{YvanTh}
\begin{equation}\eqname{beta=0simple}
\rho_{eq}(\tau=0)=\int_1\!{\mathcal
D}\phi\,\ketbra{N:\phi}{N:\phi};
\end{equation}
in the following Sec.\ref{sec:Simul}, we shall discuss how the
efficiency of an actual Monte Carlo simulation can be sensibly
improved by choosing a non-trivial {\em a priori} probability distribution.

Differently from what happened for the real-time evolution, the deterministic
term now resembles an imaginary-time Gross-Pitaevskii equation and
does not preserve any more the norm $\|\phi_{1,2}\|$ of the
wave functions. 
In particular, the deterministic contribution to the norm variation is
equal to
\begin{equation}
d\|\phi_{\alpha}\|^2=-\epsilon_{GP}[\phi_\alpha]\,\|\phi_{\alpha}\|^2\,d\tau
\eqname{DetNormeVariation}
\end{equation}
where the mean-field energy $\epsilon_{GP}[\phi]$ is defined  in terms
of the normalized wave function ${\bar \phi_\alpha}=\phi_\alpha/\|\phi_\alpha\|$ as
\begin{equation}
\epsilon_{GP}[\phi_\alpha]=\braopket{{\bar \phi}_\alpha}{h_0}{{\bar
\phi}_\alpha}+\frac{(N-1)}{2}\braopket{{\bar \phi}_\alpha{\bar \phi}_\alpha}{V_{\rm int}}{{\bar \phi}_\alpha{\bar \phi}_\alpha}.
\eqname{GPEnergy}
\end{equation}
Since interactions are assumed to be repulsive and the atoms are assumed to 
be trapped (the spectrum of $h_0$ is bounded from below)
the mean-field energy $\epsilon_{GP}$ is
also bounded from below; this 
guarantees that the stochastic equations have regular and
non-exploding solutions defined at all $\tau$'s and the stochastic
dynamics can be safely simulated by means of Monte Carlo techniques.

\section{The simulations}
\label{sec:Simul}

The mathematical manipulations of the previous section have
led to a reformulation of the interacting $N$-body problem in terms of a pair of
simple stochastic differential equations for the wave functions
$\phi_{1,2}$.
This is a good starting point for a Monte Carlo
simulation of the full quantum dynamics of the many body system:
in~\cite{GPstoch} we presented first examples of
real-time simulations, in the present paper we shall present
imaginary-time ones based on the theory of Sec.\ref{sec:Thermo}.
The present section is devoted to a general analysis of the simulation
algorithm, while the next
Sec.\ref{sec:1D} will describe the application of the method to the 
thermodynamical properties of a one-dimensional interacting Bose gas.

The simplest Monte Carlo implementation of the simulation scheme
consists in randomly choosing an initial
wave function $\phi$ on the unit sphere with an uniform distribution as
done in \eq{beta=0simple}; the two wave functions
$\phi_{1,2}$ are then evolved from their initial value $\phi_{1,2}=\phi$ according to the 
stochastic evolution \eq{EvolImagT} including the noise term 
\eq{NoiseImagT}.

Unfortunately, this simple procedure is not efficient to calculate the
expectation value of the observables at a given $\beta$.
Since the largest fraction of the randomly chosen wave functions
initially have a large mean-field energy \eq{GPEnergy}, their deterministic
evolution according to \eq{DetNormeVariation} gives a fast reduction
of the norm.
The explicit expression of the physically
interesting observables, \eq{r1} or \eq{C2}, contains the norm
of the  wave function raised to a large
power of the order $N$; in particular, the trace of the denominator of 
\eq{ObservMean} involves a sum of stochastic realizations of
\begin{equation}
\braket{\phi_2}{\phi_1}^N\propto \| \phi_1\|^N\,\|\phi_2\|^N.
\eqname{Trace}
\end{equation}
The contribution of most realizations to this sum is
therefore negligible so that a large fraction of the 
computational time is effectively wasted
and the convergence of the stochastic average is impractically slow.

A technique which is currently used in Monte Carlo integration to
overcome this kind of problems is the so-called {\em importance
sampling}, a general and comprehensive discussion of which can be found
in~\cite[\S 7.8]{NR}.
In our specific case, such a technique is equivalent to using the
following equivalent
representation of the identity operator instead of the simple \eq{beta=0simple}
\begin{equation}\eqname{beta=0modif}
\rho_{eq}(\tau=0)=\int_1\!P[\phi]{\mathcal
D}\phi\;\frac{\ketbra{N:\phi}{N:\phi}}{P[\phi]};
\end{equation}
the initial wave function is now chosen according to the (arbitrarily
chosen) {\em a priori} probability
distribution $P[\phi]$, while the weight of the
realization is correspondingly divided by a factor $P[\phi]$, which
amounts to a global rescaling of $\phi$.
By choosing a $P[\phi]$ concentrated on the realizations with largest
weight at ``time'' $\beta$, an effectively flatter integrand is obtained for 
the expectation values of the
observables which means a more efficient Monte Carlo
sampling.
Although different observables lead to different optimization schemes, 
a reasonable choice if we are mainly interested in few-body
observables such as spatial and momentum densities and low-order
correlation functions can be an optimization scheme taking the trace
of $\rho$ as an optimization criterion:
we wish that each dyadic $|N:\phi_1\rangle\langle N:\phi_2|$
in the simulation has approximately the same trace \eq{Trace}.

For a non-interacting gas, no stochastic noise is present in
the evolution equation \eq{EvolImagT}, so the two wave functions are
always equal $\phi_1=\phi_2$.
Let us first consider a dyadic $\sigma$ with initial value
$\ketbra{N:\phi(0)}{N:\phi(0)}/P[\phi(0)]$, where $\phi(0)$ is
distributed over the unit sphere with the probability density $P[\phi(0)]$.
After an 
imaginary-time evolution during $\tau$, the trace of $\sigma$ is given by
$\|\phi(\tau)\|^{2N}/P[\phi(0)]$ where $\phi(\tau)$ is the result of the
imaginary-time evolution starting from $\phi(0)$.
The
imaginary-time evolution in the absence of interactions is simply written in
the basis of eigenmodes of the trapping potential, that is the eigenvectors
of $h_0$. We label these eigenmodes with $k$
and we note their eigenenergy $E_k$:
\begin{equation}
\eqname{FourierImagTimeNI}
\phi_k(\tau)=\phi_k(0)e^{-\tau E_k/2};
\end{equation}
$\phi_k(\tau)$ denotes the component of the wave function $\phi(\tau)$
on the eigenmode $k$.
In this simple case, a unit trace for the dyadic $\sigma$ at ``time''
$\tau=\beta$ can be obtained for the
following choice of  $P[\phi(0)]$:
\begin{equation}
\eqname{PNI}
P[\phi(0)]=\|\phi(\beta)\|^{2N}=\left(\sum_k |\phi_k(0)|^2\,e^{-\beta E_k}\right)^N.
\end{equation}
Even if not the optimal one, this choice is able to provide good
efficiency also in the interacting case as we shall see on numerical 
examples.

In the next subsections we shall discuss two different methods to
numerically generate random initial wave functions according to the
probability distribution $P[\phi]$ of \eq{PNI}.
The first method (sec.\ref{sec:Brown}) has to be used when no
condensate is expected to be present. The second one
(sec.\ref{sec:Bogol}) is to be preferred at low temperatures when
there is a macroscopic occupation of a single mode; in this range of
parameters, in fact, this second method is able to provide a quicker and more
accurate sampling of $P[\phi]$, but it is extremely slow in the
absence of a condensate.

\subsection{Brownian motion sampling}
\label{sec:Brown}
A simple way to generate a random variable with a given probability
distribution function $P[\phi]$ consists in constructing a Brownian
motion  with a deterministic drift term $F\al{det}[\phi]$ and
a Langevin noise term $d\phi\al{st}$:
\begin{equation}
d\phi=F\al{det}[\phi]\,dt+d\phi\al{st}
\eqname{Langevin}
\end{equation}
whose equilibrium distribution function is exactly $P[\phi]$;
if the random variable is evolved for a sufficiently long time $t_{max}$,
its value will be independent from the initial value and distributed
according to $P[\phi]$.
In practical simulations, $t_{max}$ is chosen in a heuristic way so to make
results not to change if it is further increased. Note that $t$ is
here a fictitious evolution time that we take dimensionless.

Provided the force term is integrable
\begin{equation}
\eqname{Fd}
F\al{det}[\phi]=-2\nabla_{\phi^*} U[\phi]
\end{equation}
and the diffusion is constant and isotropic
\begin{eqnarray}
d\phi\al{st}_k d\phi^{*(st)}_{k'}=2\delta_{k,k'}\,dt, \eqname{Diffus1}
\\
d\phi\al{st}_k d\phi^{(st)}_{k'}=0
\eqname{Diffus2}
\end{eqnarray}
the equilibrium density of the Brownian motion \eq{Langevin} can be
shown to be equal to
\begin{equation}
\eqname{Peq}
P_{eq}[\phi]=e^{-2 U[\phi]}.
\end{equation}

In our specific case, the Brownian motion has to be confined on the
$\|\phi\|=1$ sphere: this is actually done by keeping only the
projection of both the deterministic force and the stochastic noise
along the sphere surface (i.e. orthogonal to $\phi$) and then
renormalizing the wave function to unity after each time step;
the renormalization procedure is equivalent to adding a deterministic
force along $\phi$ which keeps the wave function on the sphere.

Explicitly, the deterministic and stochastic increments can be
written in terms of the the projector ${\mathcal
Q}^{\perp}_{\phi}=\textbf{1}-\frac{1}{\|\phi\|^2}\ketbra{\phi}{\phi}$
as
\begin{eqnarray}
\eqname{Orthog}
d\phi'_{st}&=&{\mathcal Q}^{\perp}_{\phi}\, d\phi\al{st} \\
F'_{det}[\phi]&=&{\mathcal Q}^{\perp}_{\phi}\,F\al{det}[\phi]
\end{eqnarray}
with the force $F\al{det}$ being given by the derivative with
respect to $\phi_k^*(0)$ of the logarithm of \eq{PNI}.
\begin{equation}
F\al{det}_k=\frac{N\,\phi_k\,e^{-\beta
E_k}}{\sum_{k'}|\phi_{k'}|^2\,e^{-\beta E_{k'}}}.
\end{equation}

In order to avoid systematic errors, an actual calculation requires a
temporal step $dt$ such that both the deterministic $F'_{det}\,dt$ and the stochastic
$d\phi'_{st}$ variations of $\phi$ are small as compared to
$\|\phi\|=1$, which respectively means
$N\,dt\ll 1$ and ${\mathcal N}\,dt\ll 1$, ${\mathcal N}$ being the
number of modes actually used in the simulation.

\subsection{Bogoliubov rejection sampling}
\label{sec:Bogol}
While the sampling method discussed in the previous section can be used for any
choice of physical parameters, in the presence of a large population
in a single mode, a different method based on a Bogoliubov sampling
can be more efficient; it is also free from arbitrary parameters
like $t_{max}$ or $dt$ of the previous Brownian motion sampling which are
sources of systematic errors.

Taking into account the wave function normalization $\|\phi\|=1$, we
can eliminate from the probability distribution \eq{PNI} the field variable
$\phi_0$ using $|\phi_0|^2=1-\sum_{k\neq0}|\phi_k|^2$ (its phase is
uniformly distributed) and derive a probability
density for the $\phi_k$'s, $k\neq0$, in terms of the $\phi_{k\neq0}$ only:
\begin{multline}
d{\mathcal N}=\int\!\frac{d\phi_0}{\pi}\left(\sum_k|\phi_k|^2 e^{-\beta
E_k}\right)^N\;\delta\left[\sum_k|\phi_k|^2-1\right]\;\prod_{k\neq0} d\phi_k=\\ =\left(1-\sum_{k\neq 0} \frac{|\phi_k|^2}{\lambda_k}\right)^N\;\Theta\left[1-\sum_{k\neq0}|\phi_k|^2\right]\;\prod_{k\neq 0} d\phi_k
\eqname{Pphi2}
\end{multline}
where the $d\phi_k$'s stand for the double integral over the real and 
the imaginary parts of $\phi_k$; the energy $E_0$ of the ground $k=0$ mode
is set to $E_0=0$.
The coefficients $\lambda_{k\neq0}$ are
defined as
\begin{equation}
\lambda_k=\left(1-e^{-\beta E_k}\right)^{-1}
\end{equation}
and satisfy $\lambda_k\geq 1$.
Up to a global multiplicative factor, the
distribution function \eq{Pphi2} can be rewritten in terms of the
reduced field $\psi_k=\phi_k/\sqrt{\lambda_k}$ as
\begin{equation}
d{\mathcal N}=\left(1-\sum_{k\neq 0}
|\psi_k|^2\right)^N\;\Theta\left[1-\sum_{k\neq 0} \lambda_k|\psi_k|^2\right]\;\prod_{k\neq 0} d\psi_k
\eqname{distrPsi}
\end{equation}
and in this form it is easily sampled: $\psi_{k\neq0}$ are generated
with a spherically symmetric probability distribution proportional  to
\footnote{Thanks to the inequality
\begin{equation}
e^{-N\sum_{k\neq0}|\psi_k|^2}\geq(1-\sum_{k\neq0}|\psi_k|^2)^N:
\end{equation}
the distribution function \eq{radialP} can be sampled by means
of the {\em rejection method}~\cite[\S 7.3]{NR} with the Gaussian as the
comparison function: this latter is sampled by standard
methods~\cite[\S 7.2]{NR}; for each realization, we then choose a
random number $z$
uniformly distributed in $[0,1]$: if
\begin{equation}
z>\frac {(1-\sum_{k\neq0}|\psi_k|^2)^N}
{e^{-N\sum_{k\neq0}|\psi_k|^2}}
\end{equation}
the realization is rejected, otherwise it is retained. The remaining
realizations are automatically distributed according to \eq{radialP}.
}
\begin{equation}
(1-\sum_{k\neq0}|\psi_k|^2)^N
\eqname{radialP}
\end{equation}
and then all the realizations which do not satisfy the $\Theta$ function
in \eq{distrPsi} are {\em rejected}~\cite{NR}, i.e. the ones for which
\begin{equation}
\eqname{Reject}
\sum_{k\neq0}\lambda_k\,|\psi_k|^2>1.
\end{equation}
At the end of this procedure, $\phi$'s are obtained which are distributed
according to the desired distribution law \eq{PNI}.

The efficiency of this method clearly depends on the fraction of
realizations which have to be rejected at some stage; the step which
strongly depends on the physical parameters is the rejection governed
by condition \eq{Reject}: the condition
\eq{Reject} can be reformulated in terms of the number $N_k=N|\phi_k|^2$ of
particles in the $k$ excited mode as
\begin{equation}
\sum_{k\neq 0} N_k > N;
\eqname{Reject2}
\end{equation}
the larger the occupation of the ground $k=0$ mode, the smaller the number of realizations rejected at this stage and thus the quicker the sampling.
In presence of a macroscopic occupation, the most
probable occupation number of the $k=0$ mode is non-zero (cfr.
Sec.\ref{sec:PopStat}) so that the
rejection occurs very rarely and the Bogoliubov sampling method is
very rapid.

\section{Application: thermodynamical properties of a 1D homogeneous
Bose gas}
\label{sec:1D}
In the previous sections we have described the principles of the
stochastic approach to the imaginary-time evolution of an interacting
Bose gas as well as the crucial points of its actual implementation in
a Monte Carlo simulation.
The present section is devoted to the application of the method
developed in the previous sections to a physically relevant system
such as $N$ interacting bosons in a one dimensional box of finite size
$L$ with periodic boundary conditions.

The one-dimensional regime is not far from reach in present experiments on
transversally confined Bose condensates~\cite{1DBEC}. In particular,
the phase fluctuation of the field when the coherence length is
smaller that the spatial extension of the gas have been recently
observed in a harmonic potential~\cite{1DFluctPhase}.

\subsection{Ideal Bose gas in a large box: quantum degeneracy}

It is a well-known fact of thermodynamics that Bose-Einstein condensation does not occur
in a one-dimensional and non-interacting homogeneous system in the
thermodynamic limit $L\rightarrow\infty$ at a constant density $n=N/L$~\cite{Huang}.
Even in the absence of Bose condensation, effects coming from
the Bose statistics are however apparent in the quantum degenerate
regime~\cite{1DClass}
\begin{equation}
T<T_{deg}=\frac{2\pi \hbar^2
n^2}{m\,k_B},
\eqname{Degeneracy}
\end{equation}
i.e. when the thermal de Broglie wavelength
\begin{equation}
\lambda_{th}=\sqrt{\frac{2\pi\hbar^2}{m\,k_B T}}
\eqname{lambdadB}
\end{equation}
is larger than the mean inter particle spacing $d=n^{-1}$;
in this case, Bose statistics strongly enhances
 population of the lowest modes so that the characteristic field
coherence length 
\begin{equation}
\ell_c=\frac{\hbar^2 n}{m\,k_B T}=\frac{n \lambda_{th}^2}{2\pi}
\eqname{CohLength}
\end{equation} 
turns out to be much longer than in the case of a non-degenerate system for
which the coherence length is equal to $\lambda_{th}/\sqrt{2\pi}$.

\begin{figure}
\cen{\psfig{figure=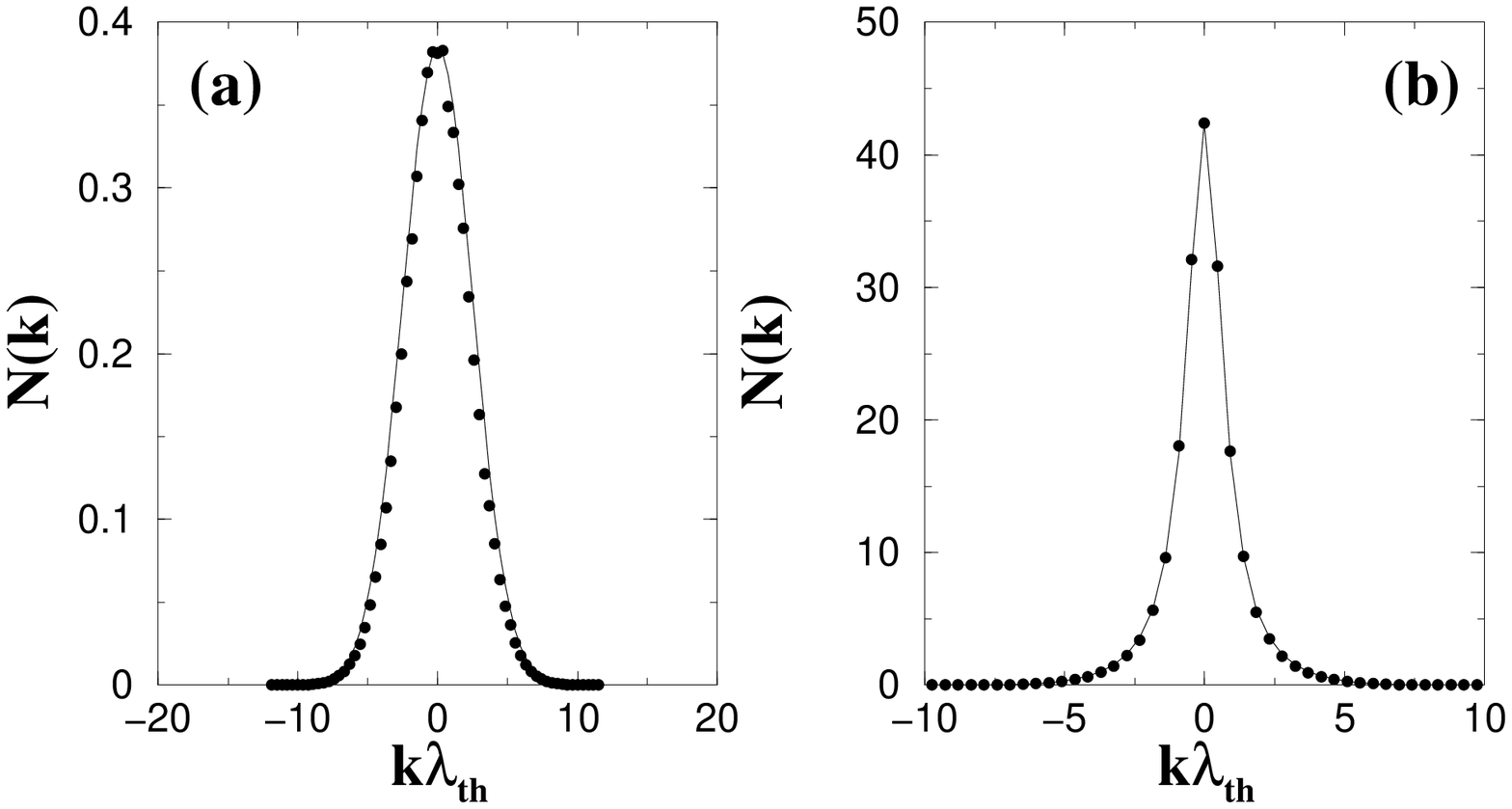,width=4in}}
\caption{
Homogeneous non-interacting gas in a large $L\gg\ell_c$ box with
periodic boundary conditions. 
Mean occupation number of the different plane wave modes of the box in the non-degenerate
($T=8\,T_{deg}$) regime (a) and in the degenerate ($T=0.02\,T_{deg}$)
regime (b); the disks are the
result of stochastic calculation with ${\mathcal N}_r=2048$ realizations, the solid line is the
prediction of grand-canonical ensemble in the thermodynamical
limit. $N=6$ atoms (a) in a box of length $L=6\,l$; $N=96$ atoms (b) in a
box of length $L=96\, l$  ($l$ is an arbitrary length unit).
\label{fig:1D}}
\end{figure}

In fig.\ref{fig:1D} the results of our stochastic
simulations for a non-interacting gas at temperatures both above and
below the degeneracy temperature are compared with the analytical predictions of the
grand-canonical ensemble. Provided the size of
the box $L$ is chosen sufficiently larger than the coherence length
$\ell_c$ (which is the case of fig.\ref{fig:1D}),
excellent agreement is found; in this limit the
canonical and the grand-canonical ensembles are in fact equivalent.
In the next section specific effects due to a fixed particle number
and the finite size of the system will be addressed.

\begin{figure}
\cen{\psfig{figure=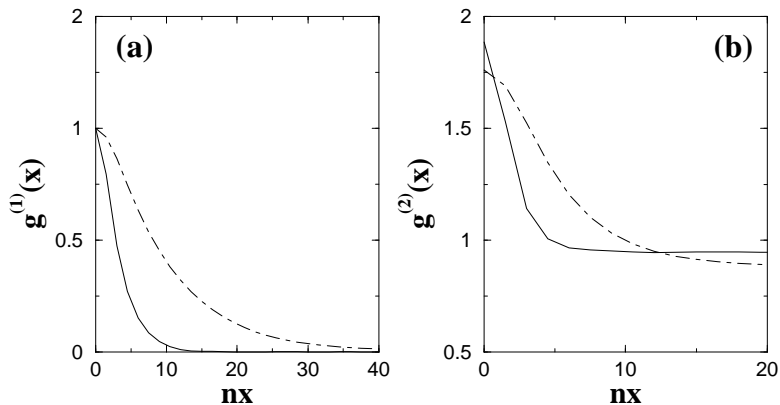,width=4in}}
\caption{\small
Homogeneous non-interacting gas in a large $L\gg\ell_c$ box. 
First (a) and second (b) order correlation functions below degeneracy
for $T=0.02\,T_{deg}$ (dot-dashed) and  $T=0.08\, T_{deg}$ (solid). 
$N=96$ atoms in a box of length $L=96\,l$ ($l$ is an
arbitrary length unit). 
Stochastic simulation with ${\mathcal N}_r=400$ realizations. 
The deviation of $g\al{2}(0)$ from the grand-canonical prediction
$g\al{2}(0)=2$ is a finite size effect (see Sec.\ref{sec:DensFluctSuppr}.)
\label{fig:1D2}}
\end{figure}

Differently from the 3D case in which a true condensate is present
in the thermodynamical limit,
the 1D first-order correlation function (fig.\ref{fig:1D2}a)
\begin{equation}
g\al{1}(x)=\frac{1}{n}\langle\Psihd(x)\Psih(0)\rangle
\eqname{g1}
\end{equation}
 has a vanishing long
distance limit $g\al{1}(x\rightarrow\infty)=0$ and a similar behavior
is found in the second-order correlation function (fig.\ref{fig:1D2}b):
\begin{equation}
g\al{2}(x)=\frac{1}{n^2}\langle\Psihd(0)\Psihd(x)\Psih(x)\Psih(0)\rangle;
\eqname{g2}
\end{equation}
in the short
distance limit, $g\al{2}$ tends to the value $2$ typical of
thermal non-interacting Bose systems; for large $x$, $g\al{2}(x)$
exponentially decays to the value $1$ with a decay constant equal to
$\ell_c/2$: the characteristic correlation length scales for the phase \eq{g1}
and density \eq{g2} fluctuations are therefore of the same order.

All the calculations presented in this section have been
performed by means of the Brownian motion sampling discussed in
Sec.\ref{sec:Brown}; in the present $L\gg\ell_c$ regime, the
efficiency of a Bogoliubov sampling would be definitely poorer.

\subsection{Ideal Bose gas in a small box: 
Bose-Einstein condensation}
\label{sec:BEC}

In the present section we shall discuss the case of a true
Bose-Einstein condensate in a finite 1D box: in particular, we shall
study the effect of a macroscopic occupation of a single mode on
different observables; differently from other techniques~\cite{QMC}, our method
is in fact able to give access to
all the observables of the system.

\subsubsection{Macroscopic occupation of a single mode}
\label{sec:MacrOcc}

In the previous section, we have shown that the phase coherence in a
1D sample extends only over a finite length $\ell_c$ and the absence of a true
condensate in the thermodynamical limit corresponds to a vanishing
limit $g\al{1}(x\rightarrow\infty)$.
On the other hand, if the sample we are considering has a length $L$
comparable to or smaller than the coherence length $\ell_c$,  phase
coherence extends over the whole sample.

According to the Wiener-Khintchine theorem~\cite{WallsMilb}, the momentum
distribution function $N(k)$ of a homogeneous system is proportional 
to the Fourier
transform of the field correlation function $g\al{1}(x)$; a phase
coherence extending over the whole sample therefore implies a
macroscopic occupation of a single mode
\footnote{Apart from a numerical factor of the order of unity, the
analogous condition for an interacting harmonically trapped cloud $\ell_c\simeq R_{TF}$ ($R_{TF}$ is the Thomas-Fermi radius
of the cloud) leads to the same condensation temperature
$T_{ph}=\frac{(\hbar \omega)^2}{\mu}N$ as predicted in~\cite{Shlyap1D}
from a different point of view.}
.

This finite-size induced Bose condensation occurs around the temperature
\begin{equation}
k_B T_{BEC}=\frac{6 N \hbar ^2}{m L^2}
\end{equation}
at which the maximum number
\footnote{
The sum can be analytically computed since the system is assumed to be
degenerate: in this case, most of the population is concentrated in
the lowest modes for which $\beta\hbar^2 k^2/2m\ll 1$, so that the
exponential can be linearized $e^{\beta\hbar^2 k^2/2m}-1\simeq \beta\hbar^2 k^2/2m$.
}
$N_{max}$ of atoms that can be stored in the excited
modes 
\begin{equation}
N_{max}=\sum_{k\neq0} \frac{1}{e^{\beta \hbar^2 k^2/2m}-1}\simeq\frac{m k_B T L^2}{6\hbar^2}
\eqname{Nmax}
\end{equation}
equals the number $N$ of atoms present in the sample; the longer $L$,
the smaller the ``transition'' temperature $T_{BEC}$. In terms of the
phase coherence length $\ell_c$, the BEC condition $N>N_{max}$ can be 
rewritten as $L<6 \ell_c$.

\begin{figure}
\cen{\psfig{figure=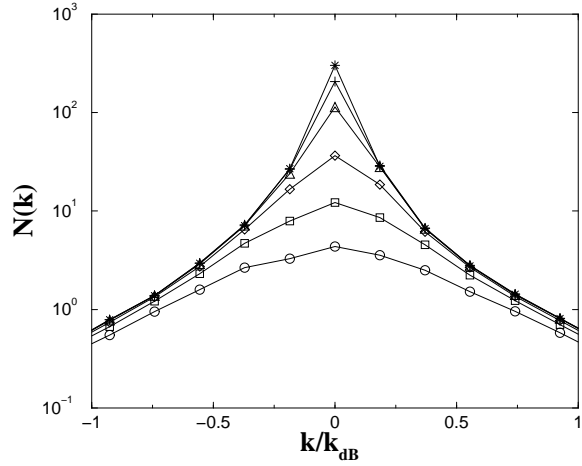,width=3in}}
\caption{\small
Degenerate non-interacting gas in a finite box:
mean occupation number of the different momentum modes 
for growing number of particles across the
Bose-Einstein condensation threshold $N/N_{BEC}=0.25$ (circles), 
$N/N_{BEC}=0.5$ (squares), $N/N_{BEC}=1$ (diamonds), $2$
(triangles), $3$ (crosses) and $4$ (stars). Stochastic calculation 
with ${\mathcal
N}_r=2048$ realizations. $L=48\,l$, $\beta=16\,\frac{m l^2}{\hbar^2}$ ($l$ is an arbitrary length unit). 
\label{fig:1DBEC1}}
\end{figure}

In fig.\ref{fig:1DBEC1} we have plotted the momentum distribution for
a growing number of atoms $N$ across the transition value $N_{BEC}$:
for $N\ll N_{BEC}$ the atoms that are further added to the system 
spread over the whole distribution.
On the other hand, for $N\gg N_{BEC}$, the population of the excited
modes saturates to the value $N_{max}$
and the added atoms accumulate in the fundamental mode; 
this behavior is strictly analogous to what happens in
three-dimensional systems.

For the three upper curves for which $N\geq 2 N_{BEC}$, the Bogoliubov
sampling technique of Sec.\ref{sec:Bogol} has revealed to be much more
efficient than the Brownian motion one;
for the lower ones, the Brownian motion technique of
Sec.\ref{sec:Brown} was instead more performing. 
As a further check, we have
successfully verified that the predictions of the two methods really
agree with each other; on the scale of fig.\ref{fig:1DBEC1} the
corresponding curves are nearly indistinguishable.

\subsubsection{Suppression of density fluctuations}
\label{sec:DensFluctSuppr}

\begin{figure}
\cen{\psfig{figure=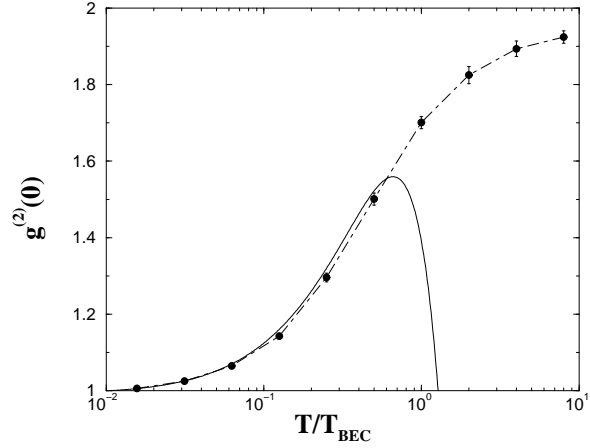,width=3in}}
\caption{\small
Degenerate non-interacting gas in a finite box: second-order
correlation function across the Bose-Einstein condensation threshold at
$T\simeq T_{BEC}$. Solid line: analytical prediction
\eq{g2final}. Disks: stochastic calculation with ${\mathcal
N}_r=256$ realizations. $N=192$ atoms in a $L=48\,l$ box  ($l$ is an arbitrary length unit).
\label{fig:1DBEC2}}
\end{figure}

In addition to the features of the momentum distribution function
and the related field correlation function $g\al{1}(x)$ discussed in
the previous section, the presence of a Bose-Einstein
condensate can be detected also by looking at the density
fluctuations quantified by the value $g\al{2}(0)$ of the second-order
correlation function \eq{g2}.
For a non Bose-condensed gas, it is well-known that $g\al{2}(0)$ is close to $2$
as for a chaotic beam of light; at zero temperature $T=0$, all the
$N\gg 1$ atoms of the system
are in the fundamental mode, so that $g\al{2}(0)=1-\frac{1}{N}\simeq 1$. As we
immediately see in fig.\ref{fig:1DBEC2}, the crossover
between the two regimes occurs in the neighborhood of the
Bose-Einstein condensation temperature $T_{BEC}$.

A simple argument allows to estimate the value of $g\al{2}(0)$ at low
temperatures, when the non-condensed fraction is small; the second
order correlation function can in fact be written as
\begin{multline}
\langle\Psihd(x)\Psihd(x)\Psih(x)\Psih(x)\rangle=\frac{1}{L^2}\left(2\sum_{k\neq
k'}\langle\ahd_k\ahd_{k'}\ah_{k'}\ah_k\rangle+\sum_k\langle\ahd_k\ahd_k\ah_k\ah_k\rangle\right)=
\\ =\frac{1}{L^2}\left(2\sum_{k\neq
k'}\langle n_k n_{k'} \rangle+\sum_k\langle n_k^2 \rangle - \sum_k\langle n_k\rangle\right).
\eqname{g2ideal}
\end{multline}
Whenever the non-condensed fraction $N_{nc}/N=\sum_k\langle n_k
\rangle/N$ is small enough, the probability for having a completely
empty condensate results negligible; under this assumption, the
non-symmetry-breeaking Bogoliubov approach of~\cite{YvanBogo} is exact
for the non-interacting gas and the spectrum of excitations is
composed of a set of harmonic oscillators, one for each $k\neq0$ mode
of the trap.
At a given temperature $T$, each mode is thermally populated in
an independent way with a mean occupation number
\begin{equation}
\langle n_k \rangle=\frac{1}{e^{\beta \hbar^2 k^2/2m}-1}
\eqname{n1}
\end{equation}
and a variance
\begin{equation}
\langle n_k^2 \rangle=2\langle n_k \rangle^2+\langle n_k \rangle.
\eqname{n2}
\end{equation}

By substituting \eq{n1} and \eq{n2} into \eq{g2ideal}, we get the
simple result
\begin{equation}
g\al{2}(0)=\left(1-\frac{1}{N}\right)\left[1+2\frac{N_{nc}}{N}\right]-\frac{N_{nc}^2}{N^2}-3\sum_{k\neq0}\frac{\langle
n_k \rangle^2}{N^2}
\eqname{g2final}
\end{equation}
which is plotted as a solid line in fig.\ref{fig:1DBEC2}.
The
agreement with the prediction of
the stochastic calculation is excellent especially at low
temperatures, when the non-condensed fraction $N_{nc}/N$ is
substantially smaller than $1$; the discrepancy becomes 
important only when the non-condensed fraction is as high as $1/2$ and
the probability of having an empty condensate mode is no longer negligible.

It is worth stressing that the present suppression of density
fluctuations is a consequence of the statistics only and therefore
completely different from the one originating from interactions that
we shall discuss in Sec.\ref{sec:1DInter}.

\subsubsection{Ground mode population statistics }
\label{sec:PopStat}

The probability of having $n$ atoms in the condensate is given by the
expectation value of the projector ${\hat Q}_{(0:n)}$ on the subspace
in which the ground $k=0$ mode contains exactly $n$ atoms:
\begin{equation}
Q_0(n)={\mathbf
Tr}[{\hat Q}_{(0:n)}\rho]\;{\mathbf
Tr}[\rho]^{-1}
\end{equation}
In terms of our Hartree ansatz \eq{Ansatz}, this quantity can be
written as 
\begin{equation}
{\mathbf Tr}[{\hat Q}_{(0:n)}\rho]=\frac{N!}{n!\,(N-n)!}\;\mean{(\phi_2^{o*}\phi_1^o)^n\;\braket{\phi_2^\perp}{\phi_1^\perp}^{N-n}},
\end{equation}
where $\phi_{1,2}^o$ are the components of the wave functions
$\phi_{1,2}$ along the ground $k=0$ mode and $\phi_{1,2}^\perp$ are
the components along the orthogonal subspace spanned by the remaining
$k\neq 0$ modes.
In the non-interacting case in which $\phi_1=\phi_2=\phi$, this
expression can be further simplified to
\begin{equation}
{\mathbf Tr}[{\hat Q}_{(0:n)}\rho]=\frac{N!}{n!\,(N-n)!}\;\mean{|\phi^o|^{2n}\;(\|\phi\|^2-|\phi^o|^2)^{N-n}}.
\end{equation}

\begin{figure}
\cen{\psfig{figure=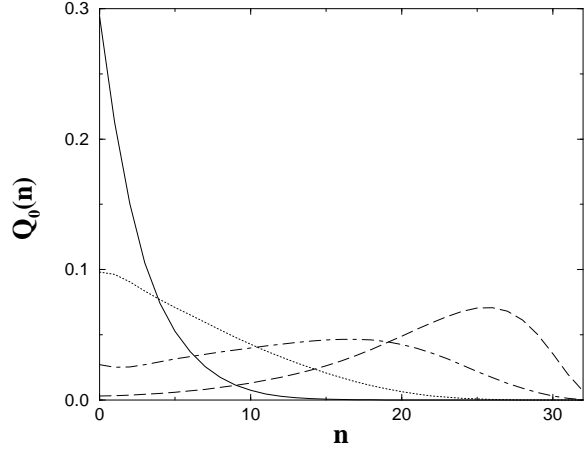,width=3in}}
\caption{\small
Non-interacting Bose gas: probability distribution of the 
ground mode occupation number
for different temperatures across the Bose-Einstein transition:
$T/T_{BEC}=30$ (solid), $3$ (dotted), $1$ (dot-dashed), $0.5$ (dashed).
Stochastic calculation with ${\mathcal N}_r=1024$ realizations. $N=32$
atoms in a $L=24\,l$ box  ($l$ is an arbitrary length unit).
\label{fig:1DBECStat}}
\end{figure}

At temperatures $T$ much higher than the condensation temperature
$T_{BEC}$, the mean population of all the modes is only a small
fraction of the total number of particles so that each of them sees
all the others as a sort of particle reservoir and its
occupation is accurately described within the
grand-canonical ensemble by the thermal distribution function 
\begin{equation}
Q_k(n)=\frac{{\bar n}^n}{(1+{\bar n})^{n+1}}
\eqname{Thermal}
\end{equation}
of average occupation number ${\bar n}$.
In this regime, the occupation probability $Q_0(n)$ of the ground mode
$k=0$ is a  
monotonically decreasing function and the most probable occupation number is $n=0$ 
(fig.\ref{fig:1DBECStat}, solid line).

As soon as $T$ approaches the condensation temperature $T_{BEC}$, the
occupation statistics of the fundamental mode is drastically
modified by the presence of a condensate (dotted and dot-dashed
lines) and, in particular its slope at $n=0$ changes its sign around 
$T=T_{BEC}$~\cite{WilkensTc}. 
For $T<T_{BEC}$ the occupation function $Q_0(n)$ has a
peaked shape around its mean value (dashed line).

A similar transition is well known to occur for the photon
distribution in a laser: below threshold the distribution has a
thermal shape \eq{Thermal}, while above threshold it tends to the
Poisson distribution typical of a coherent state~\cite{WallsMilb}.

\subsection{Interacting gas}

In the case of the non-interacting system described in the previous
section it has been possible to find the simple {\em a priori}
distribution function $P[\phi]$ \eq{PNI} which optimizes the
efficiency of the Monte Carlo calculation at a given $\beta$. 
For an interacting system, such a optimum choice is no longer simple
to implement; heuristically, the simulations of the interacting
system described in the present section have been carried
out using the same {\em a priori} $P[\phi]$ as used for non-interacting gases. 

For simplicity of the analysis, we shall limit ourselves to the case
of a zero-range interaction potential
\begin{equation}
\eqname{ZeroRange}
V_{\rm int}(x-x')=g\;\delta(x-x');
\end{equation}
the coupling constant $g$ is related to the 1D scattering length
$a_{1D}$ by
\begin{equation}
\eqname{1DScLe}
g=\frac{\hbar^2}{m a_{1D}}.
\end{equation}
On the discrete lattice we are actually using in the simulation, the delta
interaction potential \eq{ZeroRange} translates into
\begin{equation}
\eqname{ZeroRangeLatt}
V_{\rm int}(x-x')=\frac{g}{\Delta x}\;\delta_{x,x'},
\end{equation}
$\Delta x$ being the lattice spacing.

\subsubsection{Classical field regime}
\label{sec:1DInter}

If the gas is well in the degenerate regime $T\ll T_{deg}$ but not
condensed, the main
effect of atom-atom interactions is the suppression of density
 fluctuations and the appearance of a new length scale
$\xi=\sqrt{\hbar^2/2mng}$ over which the density fluctuations are
correlated~\cite{1DClass}, the so-called healing length.

\begin{figure}
\cen{\psfig{figure=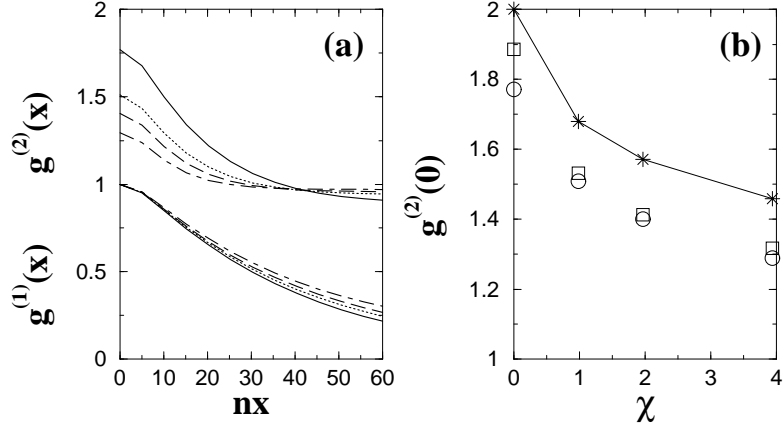,width=4in}}
\caption{\small
Left panel: first $g\al{1}(x)$ and second-order
$g\al{2}(x)$ correlation functions for different values of the
interaction strength $g/(\hbar^2/m l)=0$ (solid), $0.005$ (dotted),
$0.01$ (long-dashed) and $0.02$ (dot-dashed) ($\chi=\hbar^2 g\beta^2
n^3/m=0,1,2,4$), all within the classical field regime ($gn/k_B
T\leq 0.1\ll 1$).
Right panel: comparison of the approximate classical field
predictions in the thermodynamical limit with stochastic
simulations for finite boxes of growing size $L/\ell_c=9$ (circles), $18$
(squares). Stochastic simulation with ${\mathcal
N}_r=2048$ realizations, $\beta=0.8\,\frac{m l^2}{\hbar^2}$. $N=324$,
$L=48\,l$ (circles); $N=648$, $L=96\,l$ (squares) ($l$ is an arbitrary length unit).
\label{fig:1DInter}}
\end{figure}

If interactions are extremely weak $\chi=\hbar^2 g \beta^2 n^3/m\ll 1$, i.e. $\xi\gg\ell_c$, then the gas can be considered as an ideal gas,
the density fluctuations are close to the non-interacting value
$g\al{2}(0)=2$ and the  characteristic length scales for phase
correlation and density fluctuations turn out being of the same order
$\ell_c$ (fig.\ref{fig:1DInter}a). 
The fact that $g\al{2}(0)$ in the figure is not exactly
$2$ even in a non-interacting case is due to the fact that the system
we are considering is a finite one; we have in fact checked that for
increasing box size $L$, $g\al{2}(0)$ approaches its
thermodynamical limit of $2$ (fig.\ref{fig:1DInter}b).

For stronger interactions $\chi>1$,
density fluctuations are strongly suppressed by the interactions
and their correlation is limited to the shorter length
scale $\xi<\ell_c$; phase correlations are instead far less affected
by the presence of the interactions than density correlations
and, in particular, the field coherence length is increased by
less than a factor $2$.

From a quantitative point of view, the agreement of the result of our
stochastic simulations with the approximated predictions of classical
field theory~\cite{1DClass} looks reasonable.
We are in fact in a regime in which the classical field approximation
can not be expected to be extremely accurate, 
since the classicity parameter $\eta=g n/k_B T\leq 0.1$ and the terms
which were neglected are of order $\eta$. 

If we neglect the stochastic noise in our calculations, we are led to
an improved formulation of classical field theory: with respect to the
usual one~\cite{1DClass}, we are now including one more
``force'' term in the imaginary-time evolution of the 
wave function in addition to the classical Boltzmann weight
$e^{-\beta E[\phi]}$.
The presence of this additional term makes the approach exact at least
in the case of a non-interacting gas and, in particular, always immune
from the so-called black-body catastrophe; the new deterministic term
is in fact able to reproduce the correct Bose law even for the high
energy modes, while the Boltzmann weight alone would predict a
mean energy of $k_B T$ per mode.
In practice, we have found that the result of
the simulations is not appreciably modified if we neglect
the stochastic term provided we are well within the classicity region
$\eta\ll 1$; in other terms, this means that the predictions of the
improved classical field theory are accurate in the $\eta\ll 1$ regime.

A similar improved classical field theory can also be formulated within
the coherent state approach if we neglect the non-positive diffusion
term in the Fokker-Planck equation~\cite{1DClass}; 
also in this case, the approach gives exact predictions for the
non-interacting gas.

A characteristic feature of classical field theories based on the
Glauber-$P$ distribution such as the one in~\cite{1DClass} is that the
observables can be expressed as mean values of the corresponding
classical field variables over a real and non-negative
distribution function. 
Thanks to the Schwartz inequality, this implies that the second-order
correlation $g\al{2}(0)$ is always larger or equal to $1$ and
the field can never be antibunched.

In the next section we shall push our simulations over the boundary
$\eta=1$ of the classicity region and we shall look for signatures of
non-classical behavior such as antibunching ($g\al{2}(0)$).
In previous stochastic field approaches such as
the Glauber-$P$, such a parameter regime could not be
explored since the non-positive diffusion term could not be simulated
by means of usual Monte Carlo techniques~\cite{WallsMilb}.

\subsubsection{Beyond the classical field regime: spatial antibunching
of atoms}

In the Sec.\ref{sec:BEC} we have seen that below the Bose-Einstein
temperature $T_{BEC}$ a non-interacting gas is condensed in the
zero-momentum mode.
In the present section we discuss the effect of interactions on
a condensed sample for $\eta\geq 1$, i.e. outside the classicity region.
Since the mean occupation number of the fundamental mode is large,
 Bogoliubov rejection sampling will be used for the stochastic simulations.

\begin{figure}
\cen{\psfig{figure=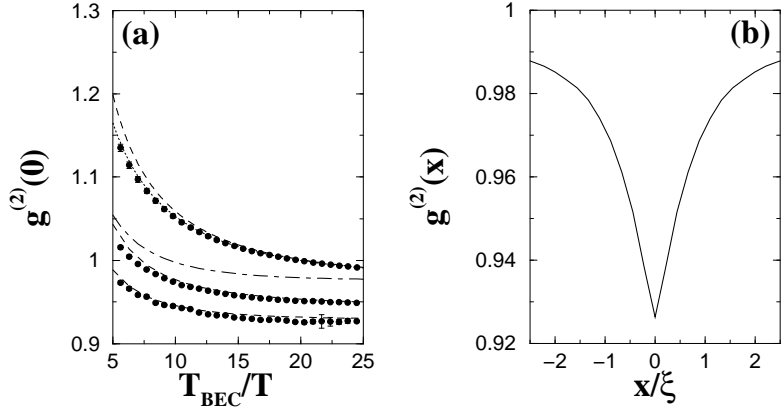,width=4in}}
\caption{\small
Left panel: second order correlation $g\al{2}(0)$ as a
function of temperature for different values of the interaction
strength; from below $\frac{m l}{\hbar^2}g=0.1, 0.05, 0$. Disks: stochastic calculation with ${\mathcal
N}_r=2048$ realizations. Dashed line: Bogoliubov prediction
(\ref{eq:g20InterAnal},\ref{eq:FreeEnergy}). Dotted line: analytical prediction \eq{g2final} for the
non-interacting gas (nearly superimposed to the disks). Dot-dashed line: stochastic calculation
neglecting the noise term for $\frac{m l}{\hbar^2}g=0.05$; notice the
low temperature limit equal to $1-1/N$.
Right panel: second order correlation function $g\al{2}(x)$ for $\frac{m l}{\hbar^2}g=0.1$
(solid) at $T_{BEC}/T=21$ obtained by means of a stochastic
calculation with ${\mathcal N}_r=2048$ realizations.
$N=42$ atoms in a $L=6\,l$ box ($l$ is an arbitrary length unit).
\label{fig:1DInterStr}}
\end{figure}

In fig.\ref{fig:1DInterStr}a we have plotted the value of $g\al{2}(0)$
as a function of temperature for different values of the interaction
constant $g$; as $T$ tends to zero, $g\al{2}(0)$ tends to a finite
value which is equal to $1-1/N$ for the non-interacting gas and equal
to a smaller value in the presence of repulsive interactions;
the stronger the interactions, the lower $g\al{2}(0)$.
In the other (b) panel, we show the behavior of $g\al{2}(x)$ well
below the condensation temperature: due to the repulsive interactions, a dip is present around $x=0$.
This effect is due to the stochastic term: the result of a calculation performed
neglecting the noise term is shown as a long-dashed line; as
expected $g\al{2}(0)$ tends to $1-\frac{1}{N}$ for $T\rightarrow
0$. The disagreement with the exact calculation 
was expected since the classical approximation 
can not be accurate in the present definitely quantum regime $\eta\simeq5$.

We now wish to compare our results for $g\al{2}(0)$ to approximate
analytical predictions. As a starting point we use the following result:
given a Hamiltonian $\Hamilt$ which depends on a parameter $g$ (in our
case, the interaction coupling constant), it follows from elementary
quantum statistical mechanics that the partial derivative of the free
energy $F=-k_B T\,\log{\mathcal Z}$ (${\mathcal Z}$ is the partition
function ${\mathcal Z}={\mathbf Tr}[e^{-\beta{\mathcal H}(g)}]$) with
respect to the interaction parameter $g$ is equal to
\begin{equation}
\frac{\partial F}{\partial g}=\expect{\frac{\partial\Hamilt}{\partial g}};
\end{equation}
in the case of a local interaction potential \eq{ZeroRange}, this
brings to the remarkable expression
\begin{equation}
g\al{2}(0)=\frac{2}{gnN}\,\frac{\partial F}{\partial g}
\eqname{g20InterAnal}
\end{equation}
which can be used to obtain an analytical prediction to be compared
with the results of stochastic simulations.

Furthermore, we are in a regime where the temperature is well below the transition temperature $T_{BEC}$
and the interactions are sufficiently weak $n\xi\gg 1$, most
of the particles are therefore in the condensate and the system is accurately
described by a Bogoliubov approximation in which the Hamiltonian of
the many-body system is approximated as a system of uncoupled harmonic
oscillators corresponding to the different elementary excitations of
the system~\cite{Bogo,Huang,CCTCollFrance,YvanBogo}.
 
In this case, the free
energy $F$ of the system of uncoupled harmonic oscillators is
immediately calculated as
\begin{equation}
\eqname{FreeEnergy}
F(g)=E_0(g)+k_B T \, \sum_{k\neq0} \log(1-e^{-\beta \epsilon_k})
\end{equation}
and its derivative with respect to $g$ has the following simple form
\begin{equation}
\frac{\partial F}{\partial g}=\frac{d E_0}{dg}+\sum_{k\neq0} \frac{d\epsilon_k}{dg}\,\frac{1}{e^{\beta\epsilon_k}-1}. 
\eqname{dFg}
\end{equation}
The index $k\neq0$ runs over the different modes whose energies are given
by the well-known Bogoliubov expression
\begin{equation}
\epsilon_k=\sqrt{\frac{\hbar^2 k^2}{2m}\left(\frac{\hbar^2 k^2}{2m}+2gn\right)}
\eqname{epsk}
\end{equation}
and the ground state energy $E_0(g)$ is equal to~\cite{YvanBogo}
\begin{equation}
E_0(g)=\frac{N(N-1)g}{2L}-\sum_k \epsilon_k |v_k|^2
\eqname{e0}
\end{equation}
with $v_k=|\textrm{sinh}\;\theta_k|^2$ and
\begin{equation}
\textrm{tanh}\;2\theta_k=\frac{gn}{\epsilon_k+gn}.
\eqname{theta}
\end{equation}
Inserting the explicit expressions (\ref{eq:epsk},\ref{eq:e0},\ref{eq:theta}) into \eq{dFg}, numerically performing the sum over
the different modes of a finite homogeneous box and finally inserting the result into \eq{g20InterAnal}
we are led to the finite temperature Bogoliubov predictions for
$g\al{2}(0)$ which are plotted in fig.\ref{fig:1DInterStr}a as dashed lines. 

These results can be compared to stochastic simulations:
for $T\rightarrow 0$, Bogoliubov theory is most accurate for weak
interactions, when the quantum depletion of the condensate is small.
At finite temperatures (but much lower that the condensation temperature),
thermal excitations are present on top of the condensate: since the
free-energy \eq{FreeEnergy} only keeps track of the lowest order terms in the
excitation density, the agreement of the approximated Bogoliubov
predictions with stochastic simulations is best at low temperatures
when the density of non-condensed atoms is small.

In the non-interacting case, the stochastic simulations can be compared also with the theory of Sec.\ref{sec:DensFluctSuppr}; as
expected, that approach is more accurate that the Bogoliubov
calculations of the present section (fig.\ref{fig:1DInterStr}a); 
in the Bogoliubov approach, in fact, the $g=0$  result is obtained as the
limit for $g\rightarrow0$ of a theory which is linearized in the excitation
density and thus able to reproduce only the terms in \eq{g2final} which
are linear in the $n_k$'s; the quadratic terms are instead not included at all.

\section{Conclusions}
\label{sec:Conclu}

In this paper, we have developed a new method for the exact
calculation of the thermodynamical quantities of an interacting
$N$-boson system in the canonical ensemble.
This method is the imaginary-time version of the one discussed in our
previous paper~\cite{GPstoch} for the real time evolution of the
system.

In order to have an efficient sampling of the observables, an
importance sampling method has been used with an {\em a priori}
distribution based on the non-interacting gas.
We have applied the simulation scheme to a homogeneous one-dimensional
interacting Bose gas and, whenever available, we have successfully
compared the predictions of stochastic simulations with the ones of
existing theories, such as the grand-canonical ensemble, the classical
field approach~\cite{1DClass} and Bogoliubov theory~\cite{Bogo,Huang,CCTCollFrance,YvanBogo}.

In particular, we have discussed the effect of Bose
condensation in finite-size systems on the different observables, such
as the momentum distribution, the density fluctuations and the
occupation statistics of the ground mode.

In the case of a
non-condensed system, the main effect of interactions is to suppress
density fluctuations without affecting in a dramatic way the phase
coherence properties of the sample.
In the classical field regime $k_B T\gg \mu$, the
effect of the noise term is negligible so that the system is already
accurately described by the deterministic force term alone.
For the opposite case of a condensed sample
at very low temperatures $k_B T\leq \mu$, interesting features
have been predicted such as an spatial antibunching of the atoms.

In the future, we plan to combine the imaginary-time evolution
discussed in the present paper with the real-time evolution discussed
in~\cite{GPstoch} in order to dispose of a tool for the numerical
calculation of dynamical properties of finite temperature Bose gases.
Improvements of the statistical properties of the stochastic simulations will hopefully be investigated both in terms of better {\em a priori}
distribution functions and in terms of the extension of the ansatz to
include more sophisticated states such as Bogoliubov vacua.
Finally, we plan to generalize the stochastic approaches in order to
consider also multi-mode nonlinear optical systems in which driving
and damping play an essential role~\cite{WallsMilb}.

The first part of the work on the real-time evolution has been done
in collaboration with Jean Dalibard, to whom we are indebted for many
stimulating discussions.
Laboratoire Kastler Brossel is a Unit\'e de Recherche de l'\'Ecole
Normale Sup\'erieure et de l'Universit\'e Paris 6, associ\'ee au CNRS.


\end{document}